\documentclass[]{spie}  

\usepackage{amsmath,amsfonts,amssymb}
\usepackage{graphicx}
\usepackage[colorlinks=true, allcolors=blue]{hyperref}

\title{The Lightspeed project: high-speed, ultra-low read noise imaging and polarimetry for the Magellan telescopes*}

\author[a,b]{Christopher Layden$^\dagger$}
\author[a,b]{Kevin Burdge}
\author[c]{John J. Piotrowski}
\author[a,h]{Gabor Furesz}
\author[a,b,d]{Juliana García-Mejía}
\author[e,f]{Jack Dinsmore}
\author[e,f]{Roger W. Romani}
\author[c]{David Osip}
\author[c,g]{Geoffrey Mo}
\author[a,b]{Vera Berger}
\author[a,i]{Aaron Householder}
\author[d]{Danielle Frostig}
\author[a,b]{Emma T. Chickles}
\author[a]{Emma Pellecer}
\author[a,b]{Malina Desai}

\affil[a]{MIT Kavli Institute for Astrophysics and Space Research, Massachusetts Institute of Technology, 77 Massachusetts Ave, Cambridge, MA 02139, USA}
\affil[b]{MIT Department of Physics, 77 Massachusetts Ave., Cambridge, MA 02139, USA}
\affil[c]{Observatories of the Carnegie Institution of Washington, 813 Santa Barbara St, Pasadena, CA 91101, USA}
\affil[d]{Center for Astrophysics \textbar\ Harvard \& Smithsonian, 60 Garden Street, Cambridge, MA 02138, USA}
\affil[e]{Department of Physics, Stanford University, Stanford, CA 94305, USA}
\affil[f]{Kavli Institute for Particle Astrophysics and Cosmology, Stanford University, Stanford, CA 94305, USA}
\affil[g]{Department of Astronomy, California Institute of Technology, 1216 E California Blvd, Pasadena, CA 91125, USA}
\affil[h]{Schmidt Sciences, Astrophysics and Space Center, New York, NY 10011, USA}
\affil[i]{Department of Earth, Atmospheric and Planetary Sciences, Massachusetts Institute of Technology, 77 Massachusetts Ave, Cambridge, MA 02139, USA}
\authorinfo{*This paper includes data gathered with the 6.5 meter Magellan Telescopes located at Las Campanas Observatory, Chile.\newline $^\dagger$Corresponding author, email: clayden7@mit.edu}

\begin{document} 
\newcommand{\LP}{\textit{Lightspeed}~}

\newcommand{\etal}{~et~al. }                    
\newcommand{\titlebar}{\rule{15cm}{1mm}\\[2.pt]\rule{15cm}{0.4mm}\\} 
\newcommand{\eg}{$e.g.$}
\newcommand{\ie}{$i.e.$}
\newcommand{\el}{$e^-$~}

\newcommand{\arcmin}{\hbox{$^\prime$}}               
\newcommand{\arcsec}{\hbox{$^{\prime\prime}$}}       
\newcommand{\kms}{ km\ s$^{-1}$}                     
\newcommand{\ergs}{erg s$^{-1}$}                     
\newcommand{\ergsM}{erg s$^{-1}$Mpc$^{-3}$}          
\newcommand{\sen}{mJy/$\sqrt{\rm Hz}$}               
\newcommand{\um}{$\upmu$m\xspace}                    
\def\degr   {$^\circ$}	
\newcommand{\ergseq}{{\rm erg}\: {\rm s}^{-1}\:}     
\newcommand{\Hub}{km\ s$^{-1}$ Mpc$^{-1}$}           

\def\K{{\rm K}}
\def\milliK{{\rm mK}}
\def\mK{{\rm \mu K}}
\def\muK{{\rm \mu K}}
\def\MJy{{\rm MJy}}
\def\Jy{{\rm Jy}}
\def\mJy{{\rm mJy}}
\def\sr{{\rm sr}}
\def\MJysr{\MJy/\sr}
\def\Mpc{{\rm Mpc}}
\def\GHz{{\rm GHz}}
\def\specint{{I_\nu}}
\def\bfzero{{\bf 0}}

\newcommand{\Msun}{$M_{\odot}$}                       
\newcommand{\Lsun}{$L_{\odot}$}                       
\newcommand{\MJ}{$M_J$}                               
\newcommand{\ME}{$M_{\oplus}$}                        
\newcommand{\Moon}{$M_{Moon}$}                        

\newcommand{\Om}{$\Omega_{\rm M}$}
\newcommand{\Ol}{$\Omega_\Lambda$}
\newcommand{\Ok}{$\Omega_k$}

\providecommand{\sorthelp}[1]{} 

\newcommand{\apj}{{\it ApJ}}
\newcommand{\apjl}{{\it ApJL}}
\newcommand{\apjs}{{\it ApJS}}
\newcommand{\aaps}{{\it A$\&$AS}}
\newcommand{\aap}{{\it A$\&$A}}
\newcommand{\aapr}{{\it A$\&$AR}}
\newcommand{\mnras}{{\it MNRAS}}
\newcommand{\aj}{{\it Astron. J.}}
\newcommand{\araa}{{\it ARAA}}
\newcommand{\nat}{{\it Nature}}
\newcommand{\pasj}{{\it PASJ}}
\newcommand{\ASP}{{\it ASP COnference Series}}
\newcommand{\CASP}{{\it Comm. Astrophys. Space Phys.}}
\newcommand{\astroph}{{\it astro-ph/}}
\newcommand{\apss}{{\item Ap\&SS}}  
\newcommand{\qjras}{{\it QJRAS}} 
\newcommand{\pasp}{{\it PASP}} 
\newcommand{\physrep}{{\it Phys. Rep.}}

\newcommand{\prl}{{\it Phys. Review Letters}}
\newcommand{\jcap}{{\it Journal of Cosmology and Astroparticle Physics}}
\newcommand\baas{{\it BAAS}}
\newcommand\pasa{{\it PASA}}
\newcommand\ssr{{\it SSRv}}

\def\aplett{{\it Astrophys.~Lett.}} 
\def\ao{{\it Appl.~Opt.}}           

\let\astap=\aap
\let\apjlett=\apjl
\let\apjsupp=\apjs
\let\applopt=\ao

\def\lsim{\mathrel{\lower2.5pt\vbox{\lineskip=0pt\baselineskip=0pt
           \hbox{$<$}\hbox{$\sim$}}}}

\def\gsim{\mathrel{\lower2.5pt\vbox{\lineskip=0pt\baselineskip=0pt
           \hbox{$>$}\hbox{$\sim$}}}}

\maketitle

\begin{abstract}
Lightspeed will be an ultra-fast ($>$\,kHz), ultra-low read noise, multicolor ($ugriz\,+$ IR) imager for the 6.5\,meter Magellan Clay telescope. In a single-channel configuration, Lightspeed will also enable single-shot linear polarimetry, narrowband imaging, or white-light imaging. Lightspeed is designed around groundbreaking single-photon-resolving detector technologies: deep sub-electron read noise CMOS image sensors from Fairchild Imaging and a HgCdTe avalanche photodiode array from Leonardo. Here we present Lightspeed's basic optical design concept, predict its on-sky performance, and highlight science cases that Lightspeed has the potential to revolutionize. Lightspeed builds upon the success of proto-Lightspeed, a single-channel prototype employing commercial-off-the-shelf re-imaging optics. proto-Lightspeed is now available as a PI instrument on the Clay telescope. We also discuss the current status of proto-Lightspeed and lessons learned from its commissioning, applicable to the development of Lightspeed and any other instrument seeking to integrate scientific CMOS image sensors.

\end{abstract}


\section{INTRODUCTION}
\label{sec:intro}
Recent rapid development in scientific complementary metal-oxide-semiconductor (CMOS) image sensors has opened new possibilities for astronomical instruments. Most naturally, CMOS image sensors delivering low read noise and fast readout could transform our ability to study astrophysical phenomena that vary on short time scales ($\lesssim 1$~s). The Lightspeed project is an effort to capitalize on this potential by integrating cutting-edge single-photon-resolving CMOS image sensors with a large telescope to perform ultra-fast imaging from the ultraviolet (UV) to the infrared (IR).

Existing high-speed optical imagers have enabled key discoveries in compact binary systems \cite{marsh:2016,kupfer:2020}, compact objects \cite{sanchez:2023,Gandhi2008,Gandhi2010,barbieri:2019,Ambrosino:2017}, outer solar system objects (via stellar occultations) \cite{morgado:2023,zhang:2023}, rapid stellar flares \cite{Kowalski:2016}, and much more. In particular, simultaneous multi-channel imagers such as ULTRACAM \cite{dhillon:2007}, HiPERCAM \cite{dhillon:2021}, and the Caltech HIgh-speed
Multi-color camERA (CHIMERA \cite{Harding2016}) (all based on charge-coupled device, or CCD, technology) deliver invaluable spectral information during these observations. Meanwhile, Geiger-mode instruments such as Iqueye \cite{naletto:2009} and Aqueye \cite{barbieri:2009} provide unparalleled temporal resolution for point sources varying at kHz timescales or faster, such as periodic optical emission from millisecond pulsars.

Today, some commercially available CMOS image sensors can nearly match the abilities of Geiger-mode devices (in terms of counting photons and delivering kHz frame rates) while maintaining the high spatial resolution and large fields offered by conventional CCDs. Such image sensors may resolve individual photons by virtue of their deep sub-electron read noise (DSERN): when read noise drops below $<0.3$~e$^-$, the user may estimate the exact number of detected photoelectrons with reasonable confidence and thereby obtain a lower ``effective" read noise \cite{ma:2022,teranishi:2012,layden:2026}. When a detector's RMS read noise drops below $\lesssim 0.15$\,e$^-$, it can deliver zero effective read noise \cite{fossum:2013} and perform perfect photon counting. At present, the most advanced commercially available DSERN CMOS image sensor is the HWK4123 from Fairchild Imaging \cite{cho:2023}, packaged in the Hamamatsu ORCA-Quest 2 camera \cite{Hamamatsu_C15550_Catalog}. This camera has been implemented in new high-speed instruments at major observatories \cite{roth:2025,lucas:2024}.

Hybridized CMOS image sensors for infrared astronomy are also rapidly advancing. In particular, HgCdTe linear mode avalanche photodiode (LmAPD) detectors can also achieve sub-electron read noise through multiplicative gain (with DSERN possible if using multiple non-destructive reads) \cite{claveau:2024,baker:2024}. Although much interest in this technology stems from its applicability to spectroscopy of Earth-like exoplanets and other faint sources \cite{claveau:2024}, the requirements for these applications ($<1\,\mathrm{e}^-$ RMS read noise per read, $<1\,\mathrm{e}^-/\mathrm{pix/ks}$ dark current) in combination with their rapid readout also make these detectors appealing for fast imaging.

Lightspeed is a planned facility instrument for the 6.5~m Magellan Clay telescope that will deliver simultaneous multicolor ($ugriz+\mathrm{IR}$) imaging using these high-speed, ultra-low read noise detectors. Lightspeed will also have a white-light channel enabling single-shot linear polarimetry. To advance the Lightspeed design and demonstrate its scientific potential, we have commissioned a single-channel prototype, proto-Lightspeed, which is now a PI instrument on the Clay telescope \cite{layden:2026}. This proceeding gives an overview of the Lightspeed project, starting with a discussion of the current status of proto-Lightspeed in Sec.~\ref{sec:protolightspeed}. We present the optical design of the full Lightspeed instrument in Sec.~\ref{sec:optomech}. In Sec.~\ref{sec:detectors}, we discuss the detectors used in proto-Lightspeed and those intended for use in Lightspeed. In Sec.~\ref{sec:performance} we estimate the performance Lightspeed will deliver, and in Sec.~\ref{sec:science} we highlight some novel science that this performance enables.

\section{PROTO-LIGHTSPEED UPDATE}
\label{sec:protolightspeed}
Since the commissioning of proto-Lightspeed in late 2025 \cite{layden:2026}, our team has completed two proto-Lightspeed observing runs to slightly improve the instrument, demonstrate its readiness for remote use by external investigators, and perform high-impact science.

\subsection{Instrument upgrades}
\label{sec:protolightspeed_upgrades}

In March 2026, we added the capability for users to move narrowband filters mounted on a rotation stage into the collimated beam of proto-Lightspeed. By rotating a narrowband filter relative to the incoming beam, the user may tune the instrument bandpass to match emission lines from galaxies at low redshifts. This upgrade extends proto-Lightspeed's deep narrowband imaging capability, which already benefits from the instrument's excellent image quality and ultra-low read noise.

We currently have two narrowband filters that may be mounted to this stage, both of size $50\times 50$\,mm$^2$. One is from Baader Planetarium with center wavelength of 671.7\,nm (S II line) and bandpass 4\,nm FWHM, while the other is from Alluxa with center wavelength of 514\,nm and bandpass 2\,nm FWHM. The center wavelength shifts blueward with a larger tilt $\theta$ from normal incidence: \[\lambda = \lambda_0 \sqrt{1-\left(\frac{n_0}{n_{eff}}\sin{\theta}\right)^2},\] with $\lambda_0$ the center wavelength at normal incidence, $n_0$ the air refractive index, and $n_{eff}$ the effective filter refractive index (typically $\lesssim 2$). The filters may be rotated up to $30^\circ$ away from normal incidence while still capturing the full collimated beam (which has diameter 36\,mm). Therefore, the 671.7\,nm filter can access the H$_\alpha$ line (rest wavelength of 656\,nm) for redshifts $0<z<0.023$, and the 514\,nm filter can access the O III line (rest wavelength of 500.7\,nm) for redshifts $0<z<0.026$. We use a Zaber RSM40B-E03T4A motorized rotation stage, which is accurate to within $\pm 0.14^\circ$. We calibrate the center wavelength as a function of filter angle using emission lines from various lamps that may illuminate the Clay telescope flat field screen.

In May 2026, we installed updated firmware to the instrument's Birger RF lens controller. As noted in \citenum{layden:2026}, the proto-Lightspeed field would shift abruptly and significantly as the instrument rotated, due to an image stabilization (IS) element of the re-imaging Canon RF lens that is not parked and locked. With the firmware update, the Birger RF controller can activate and lock the image stabilization mechanism, preventing these shifts. Even when the controller is powered off, the rotation-induced slips of the IS element are now significantly reduced (we believe because the controller when powered moved some stabilizing components in place). The magnitude of field slipping is now $\sim 5''$ peak-to-peak over a full rotation, compared to $\sim 30''$ previously.

Unfortunately, when the controller is powered and field shifts are completely eliminated, an infrared LED inside the lens body (thought to be for shutter monitoring \cite{kolarivision}) turns on, creating an enormous amount of stray light. As we have not been able to disable this LED, we installed an Astronomik L-2 IR-blocking clip filter within the Birger RF controller, eliminating proto-Lightspeed's (already poor) ability to image in the infrared. While this filter dramatically reduces the stray light level, some still leaks through, either because the LED bleeds into the visible or the filter does not provide enough suppression. The remaining light level, $\sim$1\,e$^-$/pix/s, is similar to the sky background in the $g'$ or $r'$ and is fatal for narrowband imaging. Therefore, we currently advise users to operate with the Birger RF controller unpowered and to account for small field shifts during data reduction, unless they are imaging bright sources for which the increased background light is tolerable.

\begin{figure}
    \centering
    \includegraphics[width=0.8\linewidth]{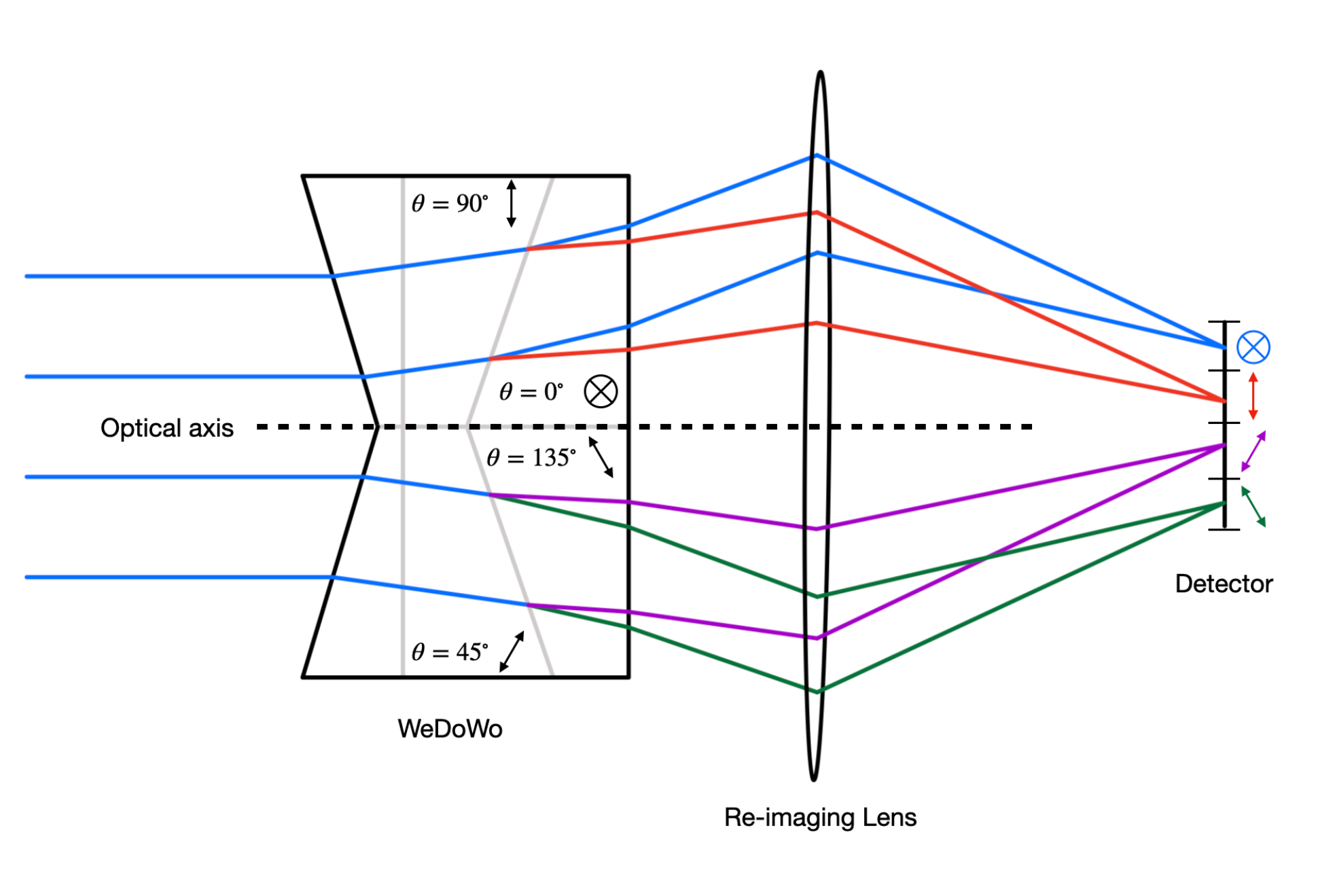}
    \caption{Operating principle of a wedged double-Wollaston (WeDoWo) prism. A WeDoWo produces four images with polarization states separated by 45$^\circ$, allowing single-shot linear polarization measurement.}
    \label{fig:wedowo}
\end{figure}

We have also designed and ordered a wedged double Wollaston (WeDoWo \cite{oliva:1997}) prism to perform single-shot linear polarimetry with proto-Lightspeed. Figure~\ref{fig:wedowo} shows the operating principle of a WeDoWo, which creates four images of a region of the field, each corresponding to a different polarization state. proto-Lightspeed will use a rectangular mask at the native telescope focal plane to define this quadruply-imaged region (of size $15'' \times 1'$) and a quarter wave plate (QWP) to measure circular polarization \cite{layden:2026}. Lightspeed will use the same strategy for polarimetry. Procurement delays have prevented us from installing the proto-Lightspeed WeDoWo and performing polarimetric calibrations, but we anticipate doing so in the fall of 2026. With proto-Lightspeed, we will develop the procedures necessary for Lightspeed to perform reliable polarimetry (which are complicated by 90$^\circ$ reflectance off the tertiary mirror), and we will determine the level of polarimetric precision both instruments can deliver.

\subsection{PI instrument status}
\label{sec:protolightspeed_status}
Beginning in 2026b, proto-Lightspeed is available to the Magellan consortium as a PI instrument, with our team supporting all observations. In our 2025b and 2026a campaigns, we extensively tested remote instrument control, identified and mitigated potential failure modes, and trained multiple team members to provide on-site support. We even identified methods to dramatically boost data transfer rates from LCO to remote observers (from $\sim 10$~Mbps to $\gtrsim 700$~Mbps), reducing the need to physically transport large data sets produced by high-speed proto-Lightspeed observations.

\subsection{Science with proto-Lightspeed}
\label{sec:protolightspeed_science}
In commissioning data alone, proto-Lightspeed delivered exciting scientific results, including optical pulsations in pulsars, optical flaring in X-ray binaries at shorter timescales than ever previously observed, and the most precise timings of eclipses in ultracompact binary systems \cite{layden:2026,berger:2026,chickles:2026}. Since then, we have used proto-Lightspeed to study more optical pulsars, X-ray binaries, and ultracompact binaries, with results forthcoming. We have also collected high-speed light curves of a diverse set of exoplanet systems (including candidate short-period transiting systems around white dwarfs and disintegrating exoplanets), performed narrowband imaging of emission line galaxies at low redshifts, and begun exploring proto-Lightspeed's capability to detect serendipitous stellar occultations by small trans-Neptunian objects.

proto-Lightspeed is limited by its COTS re-imaging optics, providing a small field of view and moderate throughput in $g'$ and $r'$ (and bad throughput elsewhere). Also, it does not deliver sufficient de-magnification for critical PSF sampling. Nonetheless, its excellent delivered image quality, ultra-low read noise, and high-speed readout have generated novel science as well as demand within the Magellan community. While we will support users of proto-Lightspeed, use it ourselves, and complete in-progress upgrades (including WeDoWo installation and continued efforts to disable the internal RF lens LED), we do not intend to make substantial changes to the fundamentally limited optical design. Instead, encouraged by the exciting scientific results and interest from the community, we are committed to advancing the design and construction of the full Lightspeed instrument that maximizes optical performance.

\section{LIGHTSPEED OPTICAL DESIGN}
\label{sec:optomech}
\begin{figure}
    \centering
    \includegraphics[width=0.98\linewidth]{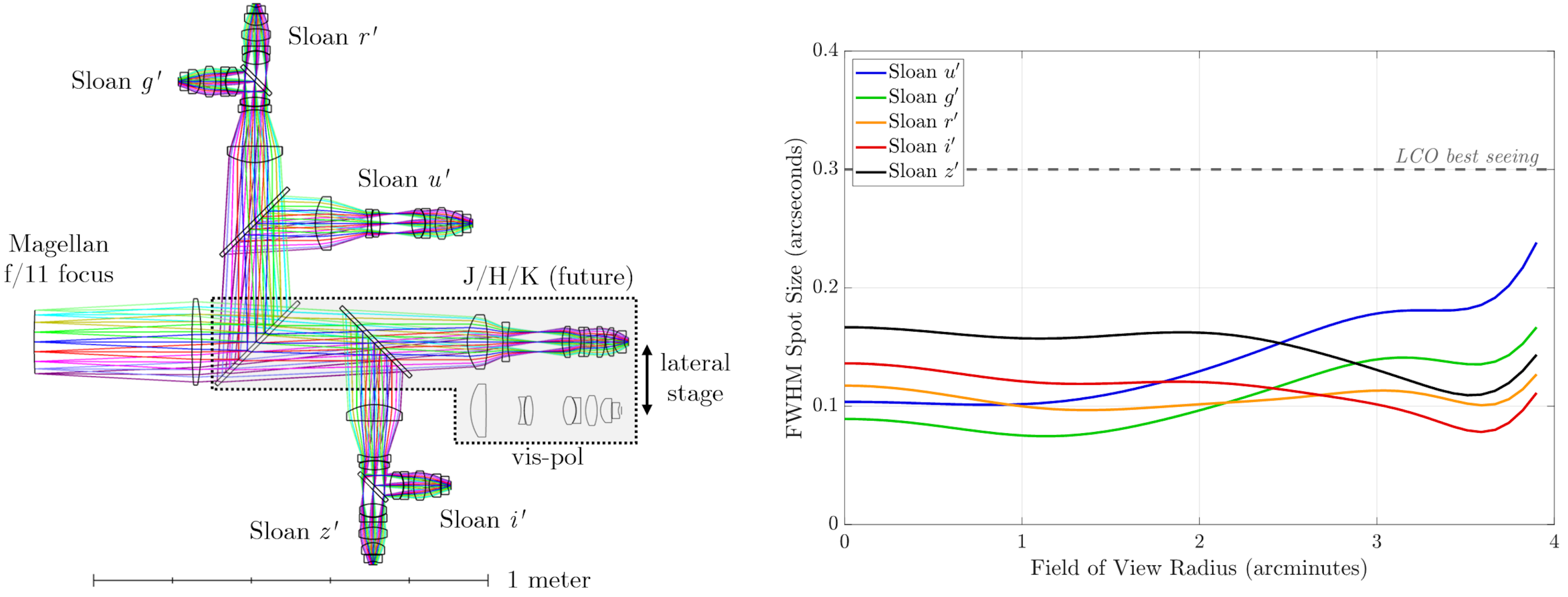}
    \caption{Left: Optical design of Lightspeed, allowing for either five-channel $ugriz$ and near-infrared simultaneous imaging or a single optical white light channel with polarimetric capabilities. Right: Designed image quality of Lightspeed's $ugriz$ channels.}
    \label{fig:optical_design}
\end{figure}

Lightspeed will combine ultra-fast, DSERN imaging with the simultaneous multicolor capabilities of highly productive instruments like CHIMERA \cite{Harding2016}, ULTRACAM \cite{dhillon:2007}, and HiPERCAM \cite{dhillon:2021}, opening up an entirely new suite of science applications. The left panel of Fig.~\ref{fig:optical_design} shows the optical design for the full Lightspeed instrument. Like CHIMERA, Lightspeed will employ dichroic beamsplitters to split incoming light into $u'$, $g'$, $r'$, $i'$, and $z'$ beams, imaging each channel simultaneously onto DSERN CMOS detectors. Unlike HiPERCAM, Lightspeed will also feature a deployable single infrared arm, enabling imaging in the $J$, $H$, and possibly $K$ bands. This IR channel can be constructed asynchronously with the $ugriz$ channels, permitting significant flexibility during construction and commissioning. Additionally, Lightspeed will have a mode of operation---referred to as the VIS-POL arm---that images optical white light ($ugriz$) onto a single detector. The VIS-POL mode produces the highest sensitivity imaging with open, wide-band and narrowband filters. For polarimetry, the pupil relief of the collimated beam is elongated to fit a WeDoWo and a low dispersion optic for spectropolarimetry.

A lateral stage holding the IR channel, two dichroic beamsplitters, and the VIS-POL channel will allow fast switching between multicolor and VIS-POL modes. Lightspeed is currently in the preliminary design phase with a majority of the work focused on the optomechanical design. We expect to begin procurement for two or more of Lightspeed's $ugriz$ channels by the end of 2026. Lightspeed will be installed on a folded port of the Magellan Clay telescope.

Lightspeed will re-image the Clay telescope's native $f/11$ focus to $f/1.4$ for the $ugriz$ and VIS-POL channels. This provides an unbinned pixel scale of $0.147''$/pix and a FOV of $7'\times7'$, greater than existing high-speed imagers (e.g., ULTRACAM \& HiPERCAM) on large telescopes ($>$3 meter aperture). The right panel of Fig.~\ref{fig:optical_design} demonstrates the image quality that will be delivered in Lightspeed's $ugriz$ channels in the central $8'$ diameter of the FOV. Lightspeed is seeing-limited across the entire wavelength range and field of view. The deployable WeDoWo in the VIS-POL channel will generate four images covering at least $3'\times0.5'$ on-sky, allowing for single-shot linear polarization measurements of extended sources.

\section{LIGHTSPEED DETECTORS}
\label{sec:detectors}
proto-Lightspeed and Lightspeed aim to capitalize on the recent development of large-format, rapid-readout, single-photon-resolving image sensors. Table~\ref{tab:detector_specs} outlines the specifications of the detectors used or considered for use in the Lightspeed project.

proto-Lightspeed uses the ORCA-Quest~2 camera \cite{Hamamatsu_C15550_Catalog}, which houses the HWK4123 CMOS image sensor from Fairchild Imaging. This camera has 2304$\times$4096 pixels of size 4.6~µm, frame rates up to 120~Hz full-frame (up to 20~kHz when windowed to 4 rows), and an RMS read noise below 0.3\,e$^-$ \cite{layden:2026,cho:2023}. We empirically found that a custom photon number resolving algorithm delivered an effective read noise $\lesssim 0.2$~e$^-$ in read noise-limited proto-Lightspeed observations \cite{layden:2026}. However, the detector provides nonlinear response at low flux levels, which is thought to result from significantly degraded internal quantum efficiency in the few-photon regime \cite{layden:2026}. Nonetheless, the ORCA-Quest~2's ultra-low read noise and kHz imaging capabilities have already enabled proto-Lightspeed users to perform novel science \cite{layden:2026,chickles:2026,berger:2026}.

Lightspeed's $ugriz$ channels are each designed to use a detector from Fairchild Imaging's next generation of DSERN CMOS image sensors. This image sensor, referred to as the STK30, offers many advantages for Lightspeed over the HWK4123. Its larger pixels (6.5~µm) and square sensor format relieve the cost and complexity required for Lightspeed's re-imaging optics to deliver a critically sampled PSF across the full detector. Its deeper depletion depth allows for good quantum efficiency in the $i'$ and $z'$ bands (estimated to be 70\% at 850~nm and 40\% at 940~nm). It is expected to deliver lower read noise ($\lesssim 0.2\,\mathrm{e}^-$ RMS), enabling an even lower effective read noise in nearly all of its pixels and true photon counting in most of them. The pixel design and operation is also expected to mitigate the type of nonlinearity that was observed in the HWK4123 detector, which will significantly boost instrument sensitivity for faint targets. 

Finally, the STK30 will provide two capabilities that are unavailable in nearly all other scientific CMOS image sensors: 2$\times$2 pixel charge summing and configurable simultaneous readout of multiple regions of interest (ROIs). In CCDs, when single-pixel spatial resolution is not required, pixel binning is a powerful technique for reducing read noise, increasing maximum frame rates, and decreasing data rates. In almost all CMOS image sensors, there is no mechanism to intentionally combine the charge in neighboring pixels. Thus, any pixel binning would be performed after charge readout and digitization, improving only data rates. In the STK30, charge can be summed in 2$\times$2 pixel bins \textit{before} readout, further decreasing read noise and increasing allowable frame rates (the full frame readout rate of 1329~Hz in Table~\ref{tab:detector_specs} assumes 2$\times$2 binning). This can be effective when atmospheric seeing is poor, such that 2$\times$2 pixel bins (0.3''$\times$0.3'') provide adequate sampling. With multi-ROI readout, a user may image small regions around targets without needing to record data in the rows or columns between the targets, thereby increasing the maximum frame rate and decreasing the data rate. This strategy will be valuable when a Lightspeed user wishes to observe a target at very high frame rates alongside spatially separated comparison stars.

We expect the STK30 to be available for procurement in late 2027 or early 2028, sufficient for Lightspeed's construction timeline. Should delays arise in STK30 procurement or should funding rapidly become available for one or more Lightspeed channels, we will temporarily employ ORCA-Quest~2 cameras for these channels.

Lightspeed's infrared channel will use a HgCdTe linear-mode avalanche photodiode (LmAPD) array detector. Such LmAPDs use avalanche gain to achieve sub-electron read noise. With high enough avalanche gain or with the combination of multiple non-destructive readout, they can deliver DSERN and thereby single-photon resolution. However, due to the excess noise factor of $\sim 1.2$ that arises from this multiplicative gain \cite{claveau:2024,CREDOne_specs}, they may not be able to reliably perform photon counting for photon numbers above unity. At present, the design assumes the use of the commercial-off-the-shelf C-RED One SWIR camera from Oxford Instruments \cite{CREDOne_specs}, which houses the Saphira detector from Leonardo.

This camera has significantly higher dark current and smaller pixel count than LmAPD arrays that Leonardo has recently developed \cite{baker:2024,lake:2020,claveau:2024}, but it will demonstrate the utility of achieving sub-electron readout noise at high frame rates in the infrared. We anticipate upgrading this channel as new devices become available.

\begin{table}[h!]
    \centering
    \begin{tabular}{c|c|c|c}
        Parameter & HWK4123\cite{layden:2026} & STK30$^*$ & Saphira $^\dagger$ \\
        \hline
         Pixel size (µm) & 4.6 & 6.5/13 & 24 \\
         Pixel format & 2304 $\times$ 4096 & 2976 $\times$ 2976 & 320 $\times$ 296 \\
         Max readout rate (Hz, full frame) & 120 & 1329 & 3500 \\
         Read noise (e$^-$, RMS) & 0.29 & 0.2 & $<0.5$ (80$\times$ avalanche gain)  \\
         Dark current (e$^-$/pix/s, mean) & 0.005 ($-20^\circ$~C) & 0.05 ($-20^\circ$~C) & $<80$ (80~K, 10$\times$ avalanche gain) \\
    \end{tabular}
    \vspace{0.1cm}
    \caption{Specifications of detectors used or planned for use in proto-Lightspeed and Lightspeed. proto-Lightspeed uses a single HWK4123 detector (in an ORCA-Quest~2 camera). Lightspeed is designed to use STK30 detectors for each of the $ugriz$ channels (with the HWK4123 as a back-up option) and, initially, a Saphira for the infrared channel. $^*$Values shared by Fairchild Imaging as estimated for the STK30 design. $^\dagger$Values per specifications of the C-RED One SWIR camera \cite{CREDOne_specs}.}
    \label{tab:detector_specs}
\end{table}

\section{PREDICTED LIGHTSPEED PERFORMANCE}
\label{sec:performance}
\begin{figure}
    \centering
    \includegraphics[width=0.7\linewidth]{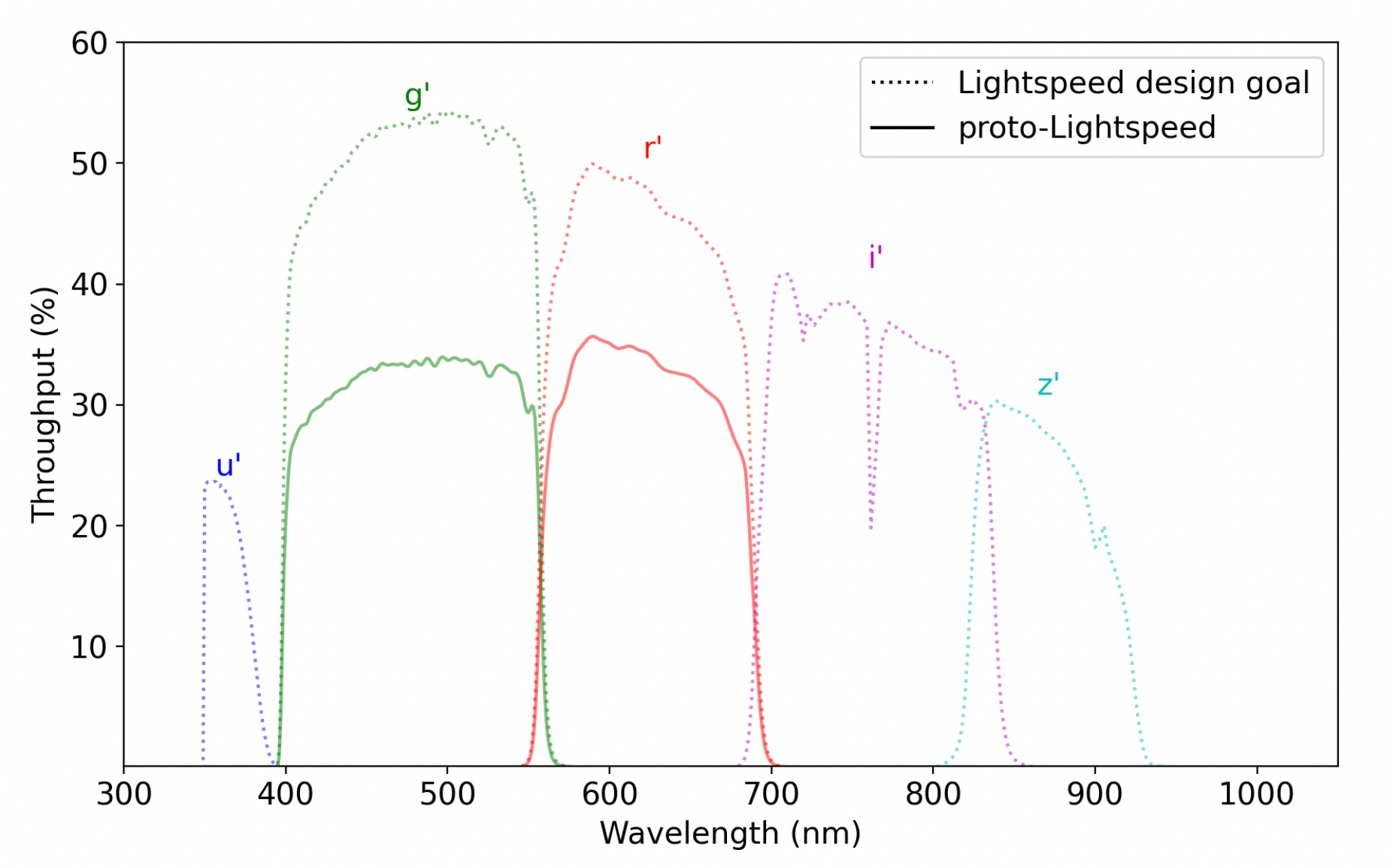}
    \caption{Design goal for Lightspeed's throughput in the $ugriz$ bands (dashed lines), compared to the throughput of proto-Lightspeed. Here, throughput accounts for atmospheric transmission, filter transmission, instruments optics transmission, and detector quantum efficiency.}
    \label{fig:lightspeed_thru}
\end{figure}

Lightspeed will be a dramatic leap forward from the in-service proto-Lightspeed---and not just because it will enable simultaneous multicolor imaging. Thanks to its custom re-imaging optics and deeper-depletion detectors, Lightspeed will deliver significantly higher throughput than proto-Lightspeed across the $ugriz$ bands. Figure~\ref{fig:lightspeed_thru} shows Lightspeed's expected throughput in these bands, compared to proto-Lightspeed's throughput in its available broad bands.

The lower read noise of the Lightspeed CMOS image sensors should enable true photon counting in most pixels. The 2$\times$2 binning option of these detectors enables optimal PSF sampling in both the best seeing conditions and in typical seeing conditions, mitigating the impact of any residual read noise. Lightspeed will employ the same GPS synchronization system as proto-Lightspeed, which has demonstrated timing stability and accuracy better than 30~µs \cite{layden:2026}.

We have developed a robust exposure time calculator (ETC) for the Lightspeed project, the properties of which are outlined in \citenum{layden:2026}. We have observed good agreement between on-sky proto-Lightspeed measurements and ETC predictions. This ETC is publicly available at \url{https://lightspeed-astro.github.io/etc.html}, and it supports performance estimates of the full Lightspeed instrument. We have used this ETC to generate the predictions for performance and scientific yield outlined in Sec.~\ref{sec:science}.

We are in the process of developing a public data reduction pipeline for proto-Lightspeed and Lightspeed. This pipeline will perform routine steps (bias/dark subtraction, flat-fielding, frame registration) as well as features bespoke to the Lightspeed project. For example, it will perform nonlinearity calibration to ensure high photometric accuracy. Most notably, it will employ a new algorithm to perform photon number resolution to the maximal extent allowable in each pixel, thereby delivering a maximum likelihood estimate of the total source flux in each frame. This method can deliver significant improvements for read noise-dominated situations, such as kHz imaging of pulsars and X-ray binaries.

\section{LIGHTSPEED SCIENCE CASES}
\label{sec:science}

\begin{figure}[t!]
    \centering
    \includegraphics[width=0.95\linewidth]{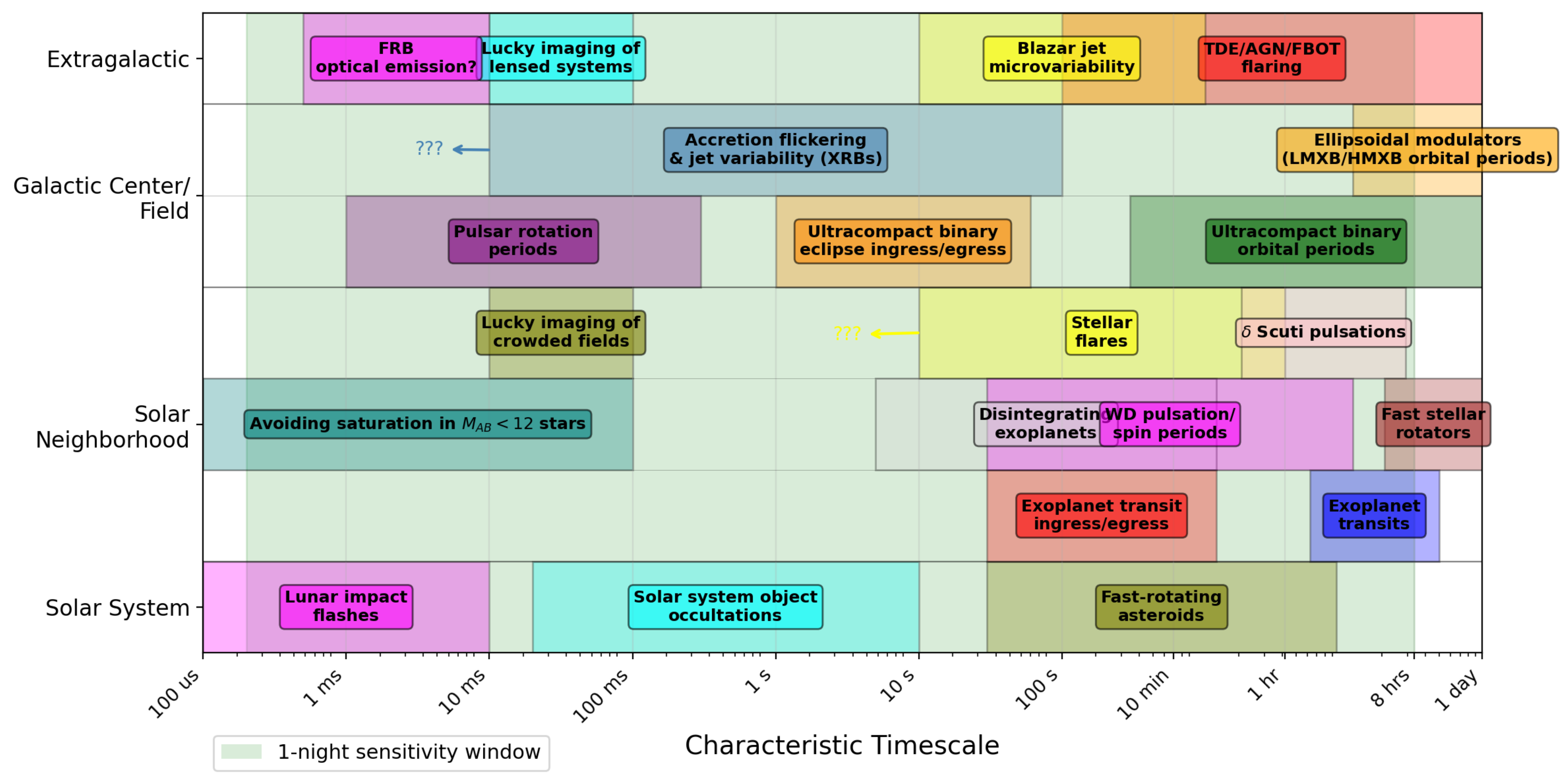}
    \caption{A sample of variable sources that Lightspeed will resolve temporally. Lightspeed's multicolor imaging and polarimetry will deliver novel insights about many of these phenomena.}
    \label{fig:science_summary}
\end{figure}

With its ultra-low readout noise, high multicolor throughput, and flexibility for polarimetry and narrowband imaging, Lightspeed can be a workhorse instrument for countless astronomical subfields. Moreover, as a facility instrument on the Clay telescope, Lightspeed could enable rapid target-of-opportunity observations of interesting transient sources. Figure~\ref{fig:science_summary} demonstrates the breadth of variable sources that Lightspeed could explore. Here we highlight just a few science cases that Lightspeed has the potential to revolutionize.

\subsection{Stellar Occultations by Solar System Objects}
Lightspeed will observe stellar occultations by a wide range of outer solar system bodies. Such bodies, including Trojans, Centaurs, Edgeworth-Kuiper Belt Objects (KBOs), and Oort Cloud objects, provide a near-pristine record of the proto-planetary disk and early solar system. By observing predicted occultations by large outer solar system bodies, we may investigate the presence and evolution of atmospheres \cite{ortiz:2012,hubbard:1993}, rings \cite{Morgado2023,santossanz:2025}, satellites \cite{sickafoose:2019}, or debris at spatial resolutions otherwise requiring in situ spacecraft missions. Lightspeed's high-SNR, multicolor light curves of such occultations will provide valuable spectral information about these features.

\begin{figure}[h!]
    \centering
    \includegraphics[width=0.98\linewidth]{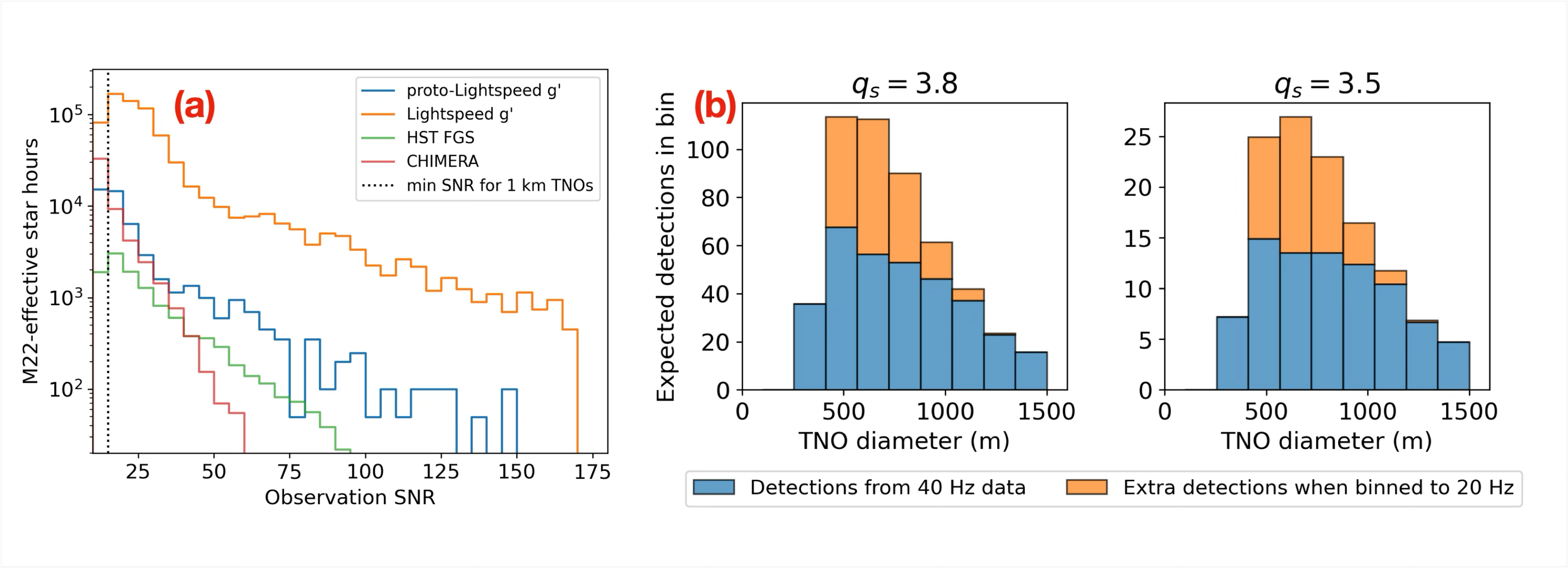}
    \caption{Depth and output of a 50-hour Lightspeed campaign for serendipitous occultations by small KBOs. (a) Expected star-hours from the Lightspeed campaign (orange) compared to an ongoing proto-Lightspeed campaign (blue) and completed campaigns using HST/FGS \cite{schlichting:2012} (green) and CHIMERA \cite{zhang:2023} (red). An ``M22-effective star hour" normalizes observation time based on ecliptic latitude and solar elongation to observations of M22 at opposition. (b) Expected yield of the Lightspeed campaign, if the cumulative distribution of small KBOs has power law index $q_s=3.8$ (as measured by \citenum{schlichting:2012}) or $q_s=3.5$. Blue bars indicate the yield from native 40\,Hz light curves. Orange bars show that by binning this data to 20\,Hz, we might identify even more KBOs.}
    \label{fig:tnos}
\end{figure}

Additionally, Lightspeed is uniquely suited to perform a deep search for serendipitous occultations by sub-km sized KBOs \cite{bickerton:2009}. Uncovering this population of small KBOs can provide valuable information about planetesimal growth and structure and the history of the outer solar system. Such objects are detectable only via occultations, with just one or two confident detections to date \cite{schlichting:2012}. To maximize the number of detected stellar occultations by small KBOs, an observing campaign should monitor many stars near the ecliptic plane. Bright globular clusters near the ecliptic are therefore good targets for such a search, and the best targets (including clusters M22, M4, and M19) pass directly overhead at LCO. The observing conditions and siting of the Magellan telescopes are ideal for this search: its well-matched latitude allows for long ($\sim 7.5$\,hr/night) observations at low airmass, its excellent image quality (typical seeing $\approx 0.6$'') reduces blending between nearby stars, and its high altitude minimizes scintillation noise. Meanwhile, Lightspeed can deliver the frame rates necessary to resolve such occultations ($\sim 40$~Hz for cold KBOs at opposition \cite{bickerton:2009}) across its $7’\times 7’$ field, with negligible read noise penalty. Lightspeed will also deliver the first multicolor light curves of km-scale TNO occultations, enabling robust false-positive rejection, precise size and distance estimates, and compositional information. A contemplated 50-hour campaign on the globular cluster M22 at opposition could increase the known number of such small TNOs by two orders of magnitude (see Fig.~\ref{fig:tnos}).

\subsection{Ultracompact binaries and orbital decay from gravitational wave emission}
\label{sec:gws}
The Zwicky Transient Facility (ZTF) has detected tens of eclipsing, ultracompact double white dwarf systems \cite{burdge:2020}, and the Vera Rubin Observatory (VRO) is expected to detect hundreds more, many of which will require dedicated high-cadence follow-up to precisely characterize their orbital periods and physical properties. These systems emit gravitational waves (GWs) detectable with the Laser Interferometer Space Antenna (LISA), making them the most numerous class of ``multi-messenger'' astrophysical sources and providing an unprecedented opportunity to combine electromagnetic and gravitational-wave observations to study compact binary evolution \cite{korol:2022, kupfer:2024}.

Lightspeed will provide outstanding electromagnetic characterization for such systems through high-speed multicolor eclipse photometry and long-term orbital-period monitoring. The instrument's sub-second cadence enables detailed measurements of eclipse ingress and egress, yielding precise constraints on binary inclination, component radii, and eclipse geometry, while simultaneous multicolor observations separate contributions of the accreting white dwarf, accretion disk, bright spot, and donor star. In the case of close double white dwarf systems, repeated eclipse timing measurements directly measure how their orbits are changing due to GW emission and mass transfer between the two objects---often competing effects \cite{chakraborty:2024}. Combined with binary parameters inferred from eclipse modeling, these measurements provide stringent tests of binary evolution models and accretion physics in interacting double white dwarf systems. They also provide essential electromagnetic context for interpreting the Galactic LISA population by breaking degeneracies in gravitational-wave parameter estimation, calibrating binary population synthesis models, and constraining the Galactic population of interacting double white dwarfs, thereby informing Type Ia supernova progenitor channels \cite{korol:2022}.

\subsection{Optical pulsars}
\label{sec:pulsars}

Periodic optical emission has been confidently detected from only a half-dozen pulsars, as only a handful of pulsars are brighter than the night sky in the optical. Of these, only two have had phase-resolved measurements of optical polarization. By providing phase-resolved colors of optical emission from young and accretion-powered pulsars, Lightspeed can connect the high energy gamma- and X-ray pulsations to the classic radio emission. By providing phase-resolved polarization measurements, Lightspeed can probe the complex and still controversial geometry of the magnetospheric emission zones. Lightspeed can even study the recently discovered optical pulsations of ``spider"-type binary pulsars \cite{Papitto_2025}, which may represent reprocessing at the inner edge of a remnant accretion disk. This provides the innermost probe of shocks in the relativistic pulsar wind at less than 10 light cylinder radii.

\begin{figure}
    \centering
    \includegraphics[width=0.98\linewidth]{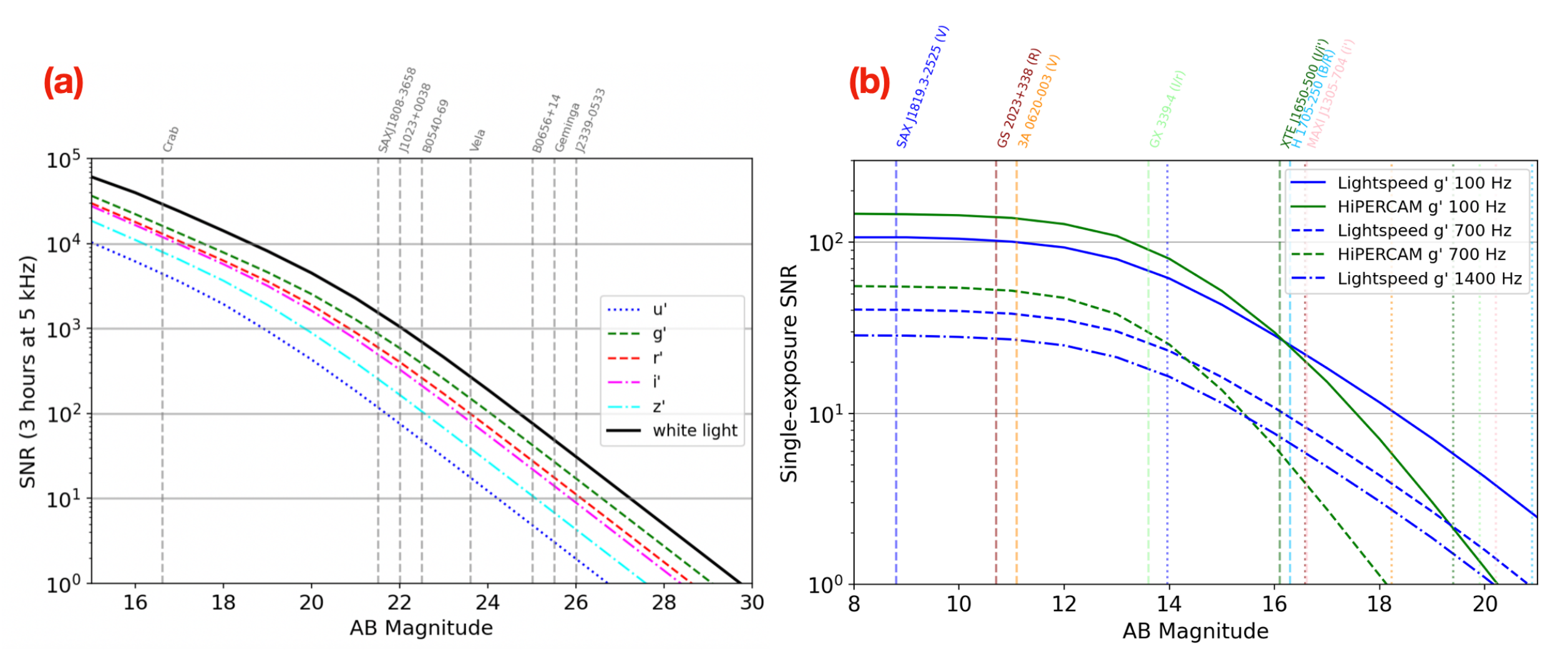}
    \caption{\textbf{(a)} Signal-to-noise ratio (SNR) of Lightspeed channels when stacking 5~kHz observations (e.g., for observations of optical pulsars). For phase-folding with $N$ bins per period, the SNR should be divided by $\sqrt{N}$. Dashed vertical lines mark $V$ magnitudes of known optical pulsars. We assume 2$\times$2 binning, 0.5'' atmospheric seeing, and 0.1~e$^-$ effective read noise (achievable through photon number resolution). \textbf{(b)} SNR achieved by the $g'$ channels of Lightspeed (blue curves) and HiPERCAM (green curves) at fast readout speeds. HiPERCAM is restricted to readout rates $<$700~Hz when using a comparison star. Vertical lines illustrate the typical apparent magnitude during outburst (dashed lines) and quiescence (dotted lines) of various dynamically confirmed BH XRBs.}
    \label{fig:pulsars_and_xrbs}
\end{figure}

With Lightspeed's imaging capabilities, we can simultaneously study the pulsar spectrum and polarization along with that of its surrounding synchrotron nebulae. Lightspeed rapid imaging also allows us to make ultra-deep searches for optically pulsed signals targeting precise positions known from radio timing, searching for pulsations well below the flux of the night sky. This can make a revolutionary advance over the few existing fast-photometry instruments which in general must search the integrated light from substantial apertures for the pulsed signal \cite{Cassanelli_2025,gradari:2011}. proto-Lightspeed has already demonstrated the utility of DSERN CMOS image sensors for these applications, as we have used it to detect optical pulsations in accreting millisecond pulsar SAX J1808.4$-$3658 at its 2.5\,ms spin period and in the 23rd-magnitude optical pulsar B0540$-$69 \cite{layden:2026}. But Lightspeed's higher throughput, improved photon counting, and pixels matched to the PSF will allow it to detect periodic optical emission from significantly fainter pulsars, as shown in Fig.~\ref{fig:pulsars_and_xrbs}. In this figure, the estimated $V$ band magnitude of accretion-powered, transitional, and redback millisecond pulsars SAXJ1808, J1023, and J2339 respectively are of the pulsed component of these systems during high states \cite{Ambrosino_2021,Illiano_2023,Papitto_2025}.

\subsection{Kilohertz imaging of X-ray binaries}
\vspace{-0.2cm}

Black hole and neutron star X-ray binaries (XRBs) sporadically undergo high-energy outbursts. The timing and energetics of optical flaring in these outbursts provide valuable insights into the geometry and emission mechanisms in these systems' extreme environments. With proto-Lightspeed, we showed that optical flaring can occur on time scales as short as 10\,ms, faster and with significantly higher peaks than previously observed with longer exposures \cite{layden:2026, berger:2026}. Resolving such rapid variability directly constrains the size of emitting regions $(r\leq c\Delta t)$, which for 10\,ms timescales corresponds to approximately an Earth radius, probing the innermost region of matter surrounding the black hole. Lightspeed will be able to conduct studies at kHz on targets as faint as 18th magnitude without being limited by read noise, meaning that it can conduct high time resolution X-ray+optical studies on the vast majority of known outbursting BH XRBs, as illustrated in Fig.~\ref{fig:pulsars_and_xrbs}.

\subsection{Exoplanets and their stars}
\label{sec:exoplanets}
Lightspeed's simultaneous multi-band photometry, flexible cadence, and photon-counting sensitivity could offer exciting new avenues to study exoplanetary systems and their host stars. One such avenue is the study of disrupted planetary material around white dwarfs: debris from tidally disrupted planetesimals produces deep, irregular, rapidly evolving transits at short orbital periods \cite{Vanderburg2015,Guidry2025}. Simultaneous multicolor, high-cadence light curves can constrain the size distribution of the dust grains comprising this debris and resolve substructure within the asymmetric transits on minute and sub-minute timescales. By monitoring these systems across multiple bands, we will aim to use Lightspeed to constrain the composition and dynamics of the accreting material, unveiling planetary end states and testing models of white dwarf pollution \cite{Brouwers2022}.

High cadence observations of `intact' planets could help us sharpen distinct exoplanet properties. In particular, planets transiting white dwarfs remain largely unexplored, with only one confirmed example to date \cite{Vanderburg2020}. Because a white dwarf is comparable in size to a terrestrial planet, a Jupiter-sized companion on a non-grazing orbit covers the entire stellar disk, producing a total eclipse lasting only minutes, with ingress and egress of order one minute. Resolving such brief and deep transits around faint white dwarfs necessitates the high cadence and ultra-low read noise of Lightspeed or similar instruments. Lightspeed also enables novel science for exoplanets orbiting main sequence stars. For example, Lightspeed could pin down mid-transit times exquisitely, enabling precise characterization of transit timing variations in multi-planet systems.





Lightspeed will also be a powerful tool for studying stellar flares, which rise in seconds and decay over minutes; extremely short events have already been observed on the nearest M dwarfs \cite{MacGregor2021}. Flares may be most consequential for low-mass M dwarfs, which host many of the nearest potentially habitable planets: their white-light and ultraviolet emission may drive the photochemistry and atmospheric escape of orbiting planets \cite{Segura2010, Tilley2019}, and their optical continuum is often used to anchor the ultraviolet portion of these stars' spectral energy distributions \cite{France2016}. A flare's inferred energy depends sensitively on how well its rapid peak is resolved: at the 2-minute and even 20-second cadences of \textit{TESS}, the peak is smeared, so energies and amplitudes may be underestimated, durations overestimated, and substructure such as quasi-periodic pulsations washed out \cite{Howard2022}. Lightspeed's sub-second sampling could capture the true peak and this substructure, while its color channels could help us measure the flare temperature and evolution \cite{Howard2020}. Together, these could yield more accurate flare energies and ultraviolet estimates, improving both the characterization of low-mass stars and assessments of their planets' radiation environments.

\section{Conclusion}
\label{sec:conclusion}
The Lightspeed project applies cutting-edge detector technologies to a large telescope to enable deep ultra-fast imaging and polarimetry. With proto-Lightspeed, we have demonstrated the potential of DSERN CMOS image sensors and made advances in studies of optical pulsars, X-ray binaries, and ultracompact binaries. We are now finalizing the design of the full Lightspeed imaging, which will deliver simultaneous multicolor imaging, with $ugriz + \textrm{IR}$ bands. The $ugriz$ channels will use Fairchild Imaging's next generation of DSERN CMOS detectors, which offer novel features for this detector class, including a large square format, RMS read noise $\lesssim 0.2\,\mathrm{e}^-$ (enabling true photon counting in most pixels), and $2\times 2$ pixel binning. The infrared channel will also be a valuable testbed for rapidly developing HgCdTe LmAPD image sensors. We intend to begin construction of at least two of Lightspeed's $ugriz$ channels by the end of this year (2026). As a facility instrument, Lightspeed might also enable early target-of-opportunity follow-up of transient sources.

\section*{Acknowledgments}
This work was funded in part by the MIT Kavli Research Investment Fund (grant MKI KRIF 2654352 and 2654373) and in part by Kavli Institute Collaboration Kickstarter (KICK) Grant PS-2025-GR-0251-3110. JGM gratefully acknowledges support from the Heising-Simons Foundation and the Pappalardo family through the MIT Pappalardo Fellowship in Physics. We warmly thank the technical support staff at Las Campanas Observatory for their assistance in machining parts, interfacing with the telescope, troubleshooting issues, and nightly operations. \cite{layden:2025}

\bibliography{report} 
\bibliographystyle{spiebib} 

\end{document}